\newif\ifarxiv
\arxivtrue          % for arXiv build
\documentclass{optica-article}

\journal{opticajournal} 
\articletype{Research Article}

\usepackage{lineno}
\ifarxiv
    \usepackage[sectionbib]{bibunits}
    \defaultbibliographystyle{unsrt}
    \defaultbibliography{references}

\else
  \linenumbers
\fi
\usepackage{graphicx}
\usepackage{dcolumn}
\usepackage{bm}
\usepackage{textcomp}
\usepackage{comment}
\usepackage{caption}
\usepackage{subcaption}
\usepackage{booktabs}
\usepackage{enumitem}
\usepackage{fancyhdr}

\usepackage[dvipsnames]{xcolor}
\usepackage{amsmath}
\usepackage{braket}
\definecolor{darkgreen}{rgb}{0,0.5,0}
\definecolor{darkblue}{rgb}{0,0,0.6}
\definecolor{purple}{rgb}{0.4,.2,0.7}

\usepackage[export]{adjustbox}
\usepackage[version=4]{mhchem}
\usepackage{siunitx}
\usepackage{hyperref}
\usepackage{cleveref}

\hypersetup{
    colorlinks=true,
    citecolor=darkgreen,
    linkcolor=black,
    urlcolor=purple
}

\begin{document}
\ifarxiv
  \begin{bibunit}
\fi

\title{\texorpdfstring{\ce{^{130}Te_2}}{130Te2} spectroscopic reference for neutral Ti lines at 391 nm and 498 nm}

\author{Matthew Bilotta,\authormark{1,2,*} Luis Castillo Gonz\'alez,\authormark{1,2} Scott Eustice,\authormark{1,2} Jackson Schrott,\authormark{1,2} and Dan M. Stamper-Kurn\authormark{1,2,3}}

\address{\authormark{1}Department of Physics, University of California, Berkeley, CA 94720\\
\authormark{2}Challenge Institute for Quantum Computation, University of California, Berkeley, CA 94720\\
\authormark{3}Materials Sciences Division, Lawrence Berkeley National Laboratory, Berkeley, CA 94720\\
Present address (M.B.): Department of Physics, Harvard University, Cambridge, MA 02138\\
Present address (L.C.G.): Universitat Aut\`onoma de Barcelona (UAB), 08193 Bellaterra, Barcelona, Spain\\
Present address (S.E.): Joint Quantum Institute, National Institutes for Standards and Technology, College Park, MD 20742}

\email{\authormark{*}mbilotta@g.harvard.edu}

\begin{abstract*}
We report on the use of ditellurium (\ce{^{130}Te_2}) as a frequency reference for laser locking at \SI{391}{\nm} and \SI{498}{\nm} optical wavelengths, which are of interest in titanium (Ti) laser-cooling experiments. In the ultraviolet region near the optical wavelength of \SI{391}{nm}, 36 previously unobserved transitions were found using laser absorption spectroscopy in a \SI{256}{GHz} range. Based on the established molecular structure of \ce{^{130}Te_2}, we attribute these lines to the $\mathrm{0_u^+\rightarrow 0_g^+}$ subsystem of the $\mathrm{^3\Sigma_u^-\rightarrow\ ^3\Sigma_g^-}$ transition with possible vibrational transitions of $\nu=(28,27,26,25,24)\rightarrow 0$ and $(27,26)\rightarrow 1$. We measure the frequencies of these lines, and also of lines near \SI{498}{\nm} wavelength, and subsequently stabilize lasers at wavelengths of \SI{391}{\nm} (and \SI{498}{\nm}) to $\sim$\SI{60}{\MHz} ($\sim$\SI{50}{\MHz}) wide resonances in \ce{^{130}Te_2}, near the optical-pumping (laser-cooling) transitions in $^{48}$Ti. We observe robust laser frequency locks, with Allan deviations of $4.9\times 10^{-10}$ ($3.6\times 10^{-11}$) at \SI{10}{s} of averaging time for the \SI{391}{\nm} (\SI{498}{\nm}) wavelength lasers.

\end{abstract*}

  %%%%%%%%%%%%%%%%%%%%%%%%%%

\section{Introduction}\label{sec:Introduction}
Molecular absorption spectroscopy is a widely used, practical route to absolute laser-frequency referencing across many areas of physics, chemistry, and optical engineering. In this approach, a laser is locked by comparing its frequency to a stable molecular transition and feeding back to a tuning actuator, yielding long-term stability without the cost and complexity of frequency combs or ultra-low expansion reference cavities. In contrast to atomic vapor references, which offer sparse resonances and can be difficult or impractical to realize with sufficient optical depth for some elements (e.g., refractory metals), molecules are furnished with dense forests of narrow lines arising from electronic, vibrational, and rotational structure, providing many convenient lock points. A few molecular species are widely used as frequency references in different spectral regions, notably iodine (\ce{I_2}), hydrogen fluoride (HF), carbon monoxide (CO), and acetylene (\ce{C_2H_2}). Extensive spectroscopic data is available for these species \cite{iodine_atlas, CO_reference, acetylene_reference, HF_reference}.

One limitation of molecular references is that they become unsuitable below a cut-off wavelength due to photodissociation of the molecules, limiting the use of such references for stabilizing visible and ultraviolet lasers. The iodine dimer, for instance, only has usable resonances at optical wavelengths longer than $\sim$\SI{500}{\nano\meter}~\cite{jun_ye_iodine}.

\par
A molecular reference known to provide transitions below the dissociation limit of iodine is ditellurium (\ce{Te_2}). In particular, \ce{Te_2} exhibits a dense set of absorption lines between its ground $\mathrm{X}$ state and the excited $\mathrm{A}$ and $\mathrm{B}$ states, spanning optical wavelengths of 381–534 nm as identified from photographic absorption spectra of hot gaseous \ce{Te_2} \cite{barrow_XB_reference}. Light in this wavelength range has myriad uses, such as laser cooling and trapping of alkaline-earth atoms and ions\cite{calcium_tellurium_423, strontium_tellurium_461, barium_ion_tellurium_455}, resonance-enhanced photoionization for isotope-selective ion-trap loading \cite{calcium_ion_trapping,ytterbium_ionization}, and fluorescence-based state detection for trapped-ion and neutral atom quantum information experiments \cite{calcium_397_state_detection, yb_flourescence}. The early film-based spectroscopy on $\mathrm{Te}_2$ provided line positions and intensities with large uncertainties, especially near the ultraviolet limit. More recently, vapor cells containing isotopically pure \ce{^{130}Te_2} have been used in precision laser spectroscopy studies, but these studies have been limited to wavelengths above $\lambda=\SI{410}{\nm}$ \cite{te2_laser_lock_again, te2_mini_spectrum}. 

\par
In this work, we carry out Doppler-broadened and Doppler-free laser absorption spectroscopy on a vapor of \ce{^{130}Te_2} across a \SI{256}{\GHz} frequency range near an optical wavelength of \SI{391}{\nm}, identifying 36 previously unobserved resonances. Motivated by the need to stabilize lasers near the optical-pumping transition in atomic titanium (Ti) \cite{titanium_MOT,eustice_optical_2023}, we determine the frequencies of the newly observed \ce{^{130}Te_2} resonances relative to the $^{48}\text{Ti}$ optical-pumping line $\mathrm{3d^24s^2\ a^3F_4}\rightarrow\mathrm{3d^2(^3P)4s4p(^3P^o)\ y^5D_4^o}$ at \SI[parse-numbers=false]{765.666836(3)_{\mathrm{stat}}(30)_{\mathrm{sys}}}{\THz} (vacuum wavelength of \SI[parse-numbers=false]{391.544257(2)_{\mathrm{stat}}(15)_{\mathrm{sys}}}{\nm}) \cite{titanium_MOT}. Additionally, in agreement with previously published spectra \cite{tellurium_atlas_original}, we identify a $^{130}\mathrm{Te}_2$ resonance located near the $^{48}\text{Ti}$ laser-cooling transition $\mathrm{3d^3(^4F)4s\ a^5F_5}\rightarrow\mathrm{3d^3(^4F)4p\ y^5G_6^o}$ at $\num{601.615883}(3)_{\mathrm{stat}}(30)_{\mathrm{sys}}$ \unit{\THz} (vacuum wavelength of \SI[parse-numbers=false]{498.312073(3)_{\mathrm{stat}}(25)_{\mathrm{sys}}}{\nm}). We measure the offsets between the $^{130}\mathrm{Te}_2$ and $^{48}\text{Ti}$ resonances by simultaneously measuring Doppler-free spectra in the $^{130}\mathrm{Te}_2$ vapor cell and a hollow cathode lamp (HCL) containing atomic Ti. Finally, using frequency modulation (FM) modulation transfer spectroscopy (MTS), we generate error signals from these $^{130}\mathrm{Te}_2$ resonances and perform proportional-integral (PI) feedback to lock the laser frequency. We then evaluate the resulting laser frequency locks, demonstrating their suitability for long-term frequency stabilization.

\section{Broadband \texorpdfstring{\ce{^{130}Te_2}}{130Te2} spectroscopy near 391 nm wavelength}\label{sec: broadband spectroscopy}

With the goal of finding resonance features at lower wavelengths than the previous blue limit of $\lambda=\SI{410}{\nm}$, we begin by performing high-resolution laser absorption spectroscopy on \ce{^{130}Te_2} at wavelengths near \SI{391}{\nm}. \Cref{fig: am optical setup} shows the layout of the optical setup. Laser light at \SI{391}{\nm} optical wavelength is produced by a Moglabs, Littrow configuration external cavity diode laser (ECDL). An isotopically pure \ce{^{130}Te_2} vapor cell rests in the center of our spectroscopy setup and is resistively heated in a furnace. The temperature is adjustable from \SI{500}{\celsius} to \SI{900}{\celsius}. At these temperatures and wavelengths, the $^{130}\mathrm{Te}_2$ vapor exhibits substantial Doppler broadening, from \SIrange{0.95}{1.2}{\giga\hertz}. To resolve the positions of resonances to a precision better than the Doppler linewidth, we employ Doppler-free spectroscopy via counter-propagating pump-probe spectroscopy. Identical spectroscopy is performed simultaneously on a Ti HCL so that all \ce{$^{130}$Te_2} resonance frequencies are measured relative to the measured $^{48}\text{Ti}$ resonance.

\label{sec:apparatus-broadband}
\begin{figure*}[t]
	\includegraphics[width=\linewidth]{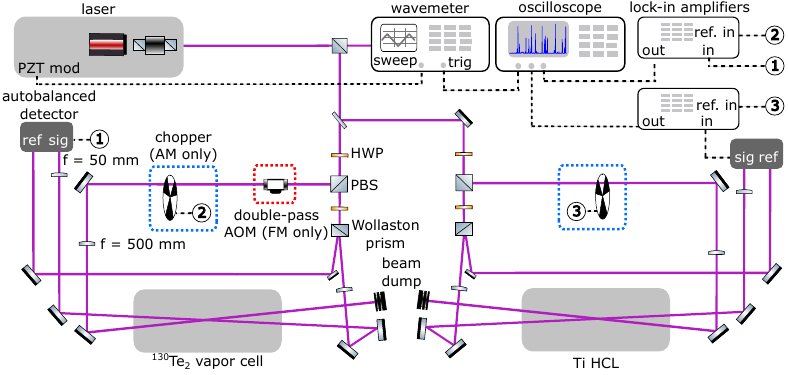}
    \caption{Experimental setup for $^{130}\mathrm{Te}_2$ and Ti spectroscopy. Light from an external-cavity diode laser is split into pump and probe arms for modulation transfer spectroscopy (MTS). The pump–probe power ratio is set with a half waveplate (HWP) and polarizing beam splitter (PBS). For amplitude modulation (AM) MTS, the probe is detected with an autobalanced receiver referenced to a pick-off beam to suppress common-mode intensity noise; the differential output is recorded during linear frequency sweeps and converted to absorbance. Moreover, the pump is chopped and the signal is demodulated with a lock-in amplifier. For frequency modulation (FM) MTS, the chopper is replaced with a double-pass acousto-optic modulator (AOM) and the autobalanced receiver is replaced with a fast photodiode.}
\label{fig: am optical setup}
\end{figure*}

To improve the signal-to-noise ratio (SNR) of the Doppler-free features, an MTS scheme is implemented: amplitude modulation (AM) is applied to the pump beam using an optical chopper at \SI{6}{\kHz}, and lock-in detection is performed on the probe beam signal \cite{MTS_paper}. Additionally, the cell temperature is tuned to optimize the SNR of the demodulated MTS signal. For spectroscopy, we choose \SI{600}{\celsius}, as it maximizes the SNR of the MTS signal at the resonance closest to the $\mathrm{a^3F_4}\rightarrow\mathrm{y^5D_4^o}$ optical-pumping transition in Ti. See appendix~\ref{sec:temp-scans} for further details on the temperature dependence of our observations.

\par

Frequency sweeps of the \SI{391}{\nm} light are performed by driving the slow cavity piezo within the ECDL. The laser frequency is determined using a High-Finesse wavemeter. We observe $\sim$\SI{3}{\MHz} drifts of our wavemeter calibration on \SI{10}{\minute} timescales, which is a systematic uncertainty in the relative frequencies of all the lines we measure in this study. The data acquisition from the wavemeter and the spectroscopy traces are synchronized to the laser scan, ensuring that the recorded signals are referenced consistently across all scans. 
The wavemeter is calibrated to a laser locked to the \ce{^{87}Rb} D$_2$ line (wavelength \SI{780}{nm}). During the individual frequency scans, the laser output power varies by roughly 50\%. To eliminate the impact of these power variations on the spectra, we measure cell transmission with an autobalanced photoreceiver, which corrects for common-mode intensity noise to the reference and signal ports. This noise-subtracted signal can then be used to determine the absorbance of the $^{130}\mathrm{Te}_2$ vapor.

To explore a large optical frequency band, we carry out a piece-wise scan by adjusting the ECDL grating between segments. In this way, we cover a \SI{256}{\GHz} span between \SI{765.552}{\THz} and \SI{765.808}{\THz}. Figure~\ref{fig: 391_large_scan} shows both Doppler-broadened and Doppler-free spectral traces across the full region. Over this span we observe 36 transitions, yielding an average density of approximately one resonance every \SI{7.1}{\GHz}, as listed in \Cref{tab:doppler_vs_MTS_sorted}. Compared to I$_2$, \ce{^{130}Te_2} exhibits fewer spectral lines due to the absence of hyperfine structure from its $I=0$ nuclear spin.

\begin{figure}[t]
	\centering
		\includegraphics[width=\linewidth]{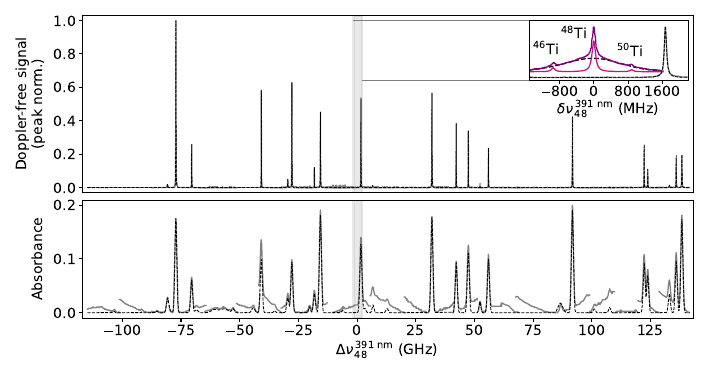}
	\caption{Doppler-broadened and saturated absorption spectrum of \ce{^{130}Te_2} (dark grey) over $256$ GHz scan at \SI{600}{\celsius}. The frequency axis is defined relative to the absolute frequency of the $\mathrm{a^3F_4}\rightarrow\mathrm{y^5D_4^o}$ line used for optical-pumping of $^{48}\text{Ti}$ at $\lambda=\SI{391}{\nm}$. The total spectrum is the result of a piece-wise concatenation of 26 different scans, resulting in $36$ total transitions (\Cref{tab:doppler_vs_MTS_sorted}). Transitions are determined by fitting each piece-wise scan to a sum of Gaussians plus a linear offset to account for an etalon. The inset shows the MTS signal of \ce{^{130}Te_2} (dark grey) near the MTS spectrum of the $\mathrm{a^3F_4}\rightarrow\mathrm{y^5D_4^o}$ optical-pumping transition of Ti at \SI{391}{\nm} (dark purple). The Ti spectrum is fit with 3 Lorentzians added atop a Gaussian background. The fit of the 3 Lorentzians (light purple) shows clear peaks for the three $I=0$ isotopes of Ti ($^{46}\text{Ti}, ^{48}\text{Ti}, ^{50}\text{Ti}$). The Gaussian background (black, dashed line) is produced by velocity-changing collisions between the HCL buffer gas and the Ti atoms excited by the pump \cite{e8_isotope_shifts_2021}. A small vertical offset is applied between the two spectra for clarity. For this work, we define absorbance as $\log_{10}(P_\mathrm{in}/P_\mathrm{out})$, where $P_\mathrm{in}$ and $P_\mathrm{out}$ are the input and output optical powers from the vapor cell, respectively.}
\label{fig: 391_large_scan}
\end{figure}

Across the same \SI{256}{GHz} range, Doppler-free signals are recorded and analyzed. We fit each absorbance trace to a sum of Gaussian functions, with the number of Gaussians chosen to match the number of resolvable transitions above the noise floor. While Doppler broadening is substantially reduced in MTS, the observed linewidths remain broadened by several mechanisms. In our data, the measured MTS full-width at half-maximum (FWHM) values range from approximately \SIrange{35}{250}{\MHz} (Table~\ref{tab:doppler_vs_MTS_sorted}). We believe the dominant contributions to this broadening are pressure broadening, which is enhanced by the high vapor-cell temperature, and residual Doppler broadening arising from a nonzero pump-probe crossing angle. For our operating conditions, the pressure-broadened contribution is estimated to be \SI{18(6)}{\MHz} FWHM \cite{pressure_vs_temp_pressure_broadening}, and the residual Doppler broadening is estimated as $\leq$\SI{29(7)}{\MHz} using $\nu_{\mathrm{rD}}=\nu_{\mathrm{D}}\sin(\theta/2)$, where $\theta$ is a conservative upper-limit measurement of the pump--probe crossing angle \cite{Residual_doppler_broadening}. By comparison, other broadening mechanisms are much smaller: the transit-time broadening is \SI{0.22}{\MHz} for a beam waist of \SI{0.6}{\mm}, and the power broadening is expected to be minor relative to the observed linewidths based roughly on previously reported \ce{^{130}Te_2} saturation intensities on similar transitions\cite{calcium_tellurium_423}. Thus, the narrowest observed MTS features are fairly consistent with the combined effect of pressure and residual Doppler broadening. Some resonances are nevertheless substantially broader than \SI{35}{\MHz}; in those cases, we cannot determine with certainty whether the observed feature corresponds to a single broad resonance or to multiple overlapping resonances.

Based on the established molecular structure, the observed features belong to the $\mathrm{0_u^+\rightarrow 0_g^+}$ subsystem of the $\mathrm{^3\Sigma_u^-\rightarrow\!\,^3\Sigma_g^-}$ transition \cite{barrow_XB_reference}. By comparing with band-head data \cite{band_head_data}, we infer that the observed lines have possible vibrational assignments of $\nu=(28,27,26,25,24)\rightarrow 0$ and $(27,26)\rightarrow 1$. The states are well described by Hund's case (c) due to strong spin–orbit coupling, with good quantum numbers $\Omega$, $J_{\mathrm{e}}$, and $J$. The upper and lower electronic states are referred to as $\mathrm{B0_u^+}$ and $\mathrm{X0_g^+}$, respectively. Uncertainties in the rovibrational molecular constants and spectral perturbations from molecular level crossings preclude a more precise rovibrational transition assignment. Further discussion of our fitting and assignment of the \ce{^{130}Te_2} lines is given in appendix~\ref{sec:line-assignment}.

This wide-range survey furnishes the frequency map needed for targeted laser stabilization. In the next section we focus on resonances near the Ti optical-pumping and laser-cooling transitions, at wavelengths of \SI{391}{nm} and \SI{498}{nm}.

%%%%%%%%%%%%%%%%%%%%%%%%%%%%%%%%%%%%%%%%%%%%%%%%%%%%%%%%%%%%%%%%%%%%%%%%%%%%%%%%%%%%%%%%%%
\section{Laser locking using \texorpdfstring{\ce{^{130}Te_2}}{130Te2} references to address Ti optical resonances}
\label{sec:identification-locking}

We now consider a specific application of the $^{130}\mathrm{Te}_2$ vapor frequency references: stabilizing lasers close in frequency to the laser cooling and optical pumping transitions in $^{48}$Ti. Toward this aim, we identify molecular transitions whose optical frequencies lie within several GHz of the relevant Ti atomic transitions---mindful that the remaining frequency offset between the $^{130}\mbox{Te}_2$ and Ti resonances can be spanned by GHz-range electro-optical modulation or offset locks---and for which we observe strong, Doppler-free molecular resonances in the MTS signal. From the spectroscopy in the preceding section, we identify a $^{130}\mathrm{Te}_2$ resonance at \SI[parse-numbers=false]{765.668495(10)_{\mathrm{stat}}(30)_{\mathrm{sys}}}{\THz} with a MTS linewidth of \qty{63.0(7)}{\MHz} that lies \SI{1.659(10)}{\GHz} above the $\mathrm{a^3F_4}\rightarrow\mathrm{y^5D_4^o}$ optical-pumping transition of Ti at \SI{391}{\nm} wavelength (see the inset of Fig.~\ref{fig: 391_large_scan} and Table~\ref{tab:doppler_vs_MTS_sorted} in appendix~\ref{sec:line-assignment}). Using existing spectra as a guide\cite{tellurium_atlas_original, te2_mini_spectrum}, we also identify a $^{130}\mathrm{Te}_2$ absorption feature in the vicinity of the Ti cooling transition at \SI{498}{\nm} wavelength. Fig.~\ref{fig: 498_tivste2} shows simultaneously measured AM MTS spectra of this line and the laser-cooling transition in Ti. From these spectra we measure a frequency offset of $\delta\nu^{\qty{498}{\nm}}_{48}=$\qty{-299.5(1)}{\MHz} between the \ce{^{130}Te_2} and \ce{^{48}Ti} lines, and an MTS linewidth of \qty{50.5(3)}{\MHz} for the \ce{^{130}Te_2} resonance.

\begin{figure}[ht!]
	\centering
		\includegraphics{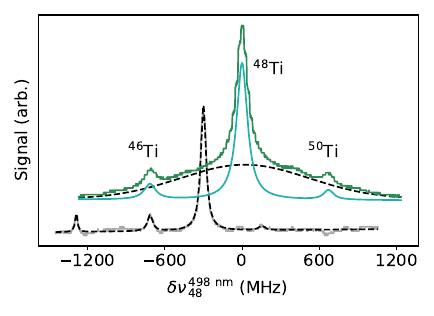}
	\caption{Amplitude-modulated spectrum of $^{130}\mathrm{Te}_2$ (light grey) at \SI{725}{\celsius} and saturated absorption spectrum of Ti (dark green) at $\lambda = \SI{498}{\nm}$. The frequency axis is defined relative to the absolute frequency of the $^{48}\text{Ti}$ $\mathrm{a^5F_5}\rightarrow\mathrm{y^5G_6^o}$ line used for cooling. The fit of 3 Lorentzians (light green) shows clear peaks for the three $I=0$ isotopes ($^{46}\text{Ti}, ^{48}\text{Ti}, ^{50}\text{Ti}$) by subtracting from the Ti signal the Gaussian background (black, dashed line) that is produced by velocity-changing collisions of atoms that are excited by the pump \cite{e8_isotope_shifts_2021}. The $^{130}\mathrm{Te}_2$ signal is fitted with 4 Lorentzians (black, dashed line). A small vertical offset is applied between the two spectra for clarity.} 
\label{fig: 498_tivste2}
\end{figure}

Having identified $^{130}\mathrm{Te}_2$ resonances near the relevant Ti transitions, we next lock the laser frequency to these molecular spectroscopy features. To generate an error signal for stabilization, we implement FM MTS by replacing the optical chopper with a double-pass AOM in the pump path (see Fig.~\ref{fig: am optical setup}). To accommodate the higher detection bandwidth required for FM operation, we also replace the autobalanced receiver with a fast photodiode while keeping the pump-probe geometry unchanged. The AOM is driven by a voltage-controlled oscillator (VCO) whose frequency is dithered at \SI{145}{\kHz}, with the modulation parameters adjusted to maximize the error-signal slope while maintaining a high signal-to-noise ratio. The double-pass configuration also provides a convenient fixed frequency offset between the lock point and the transition center.

The lock performance is quantified on short timescales by monitoring the power spectral density of the feedback error signal and on long timescales by reference to a wavemeter. At \SI{391}{\nm} wavelength, we obtain robust locks, with a long-term frequency distribution characterized by a Gaussian width of \SI{5.5}{\MHz} (Fig.~\ref{fig: allan_dev}(b)); the capture range of the error signal is approximately \SI{25.8}{\MHz} (Fig.~\ref{fig: allan_dev}(a)). Relative to the unlocked case, we found that the drift over a 4-hour span is reduced by a factor of about 30. At \SI{498}{\nm} wavelength, the $^{130}\mathrm{Te}_2$-referenced lock exhibits a Gaussian frequency width of \SI{0.2}{\MHz} (Fig.~\ref{fig: allan_dev}(b)), with an observed capture range of \SI{29.8}{\MHz} (Fig.~\ref{fig: allan_dev}(a)). These results demonstrate that the identified $^{130}\mathrm{Te}_2$ resonances provide stable, high-contrast error signals suitable for long-term operation near both Ti transitions.

\begin{figure}[t!h]
	\centering
		\includegraphics[width=9cm]{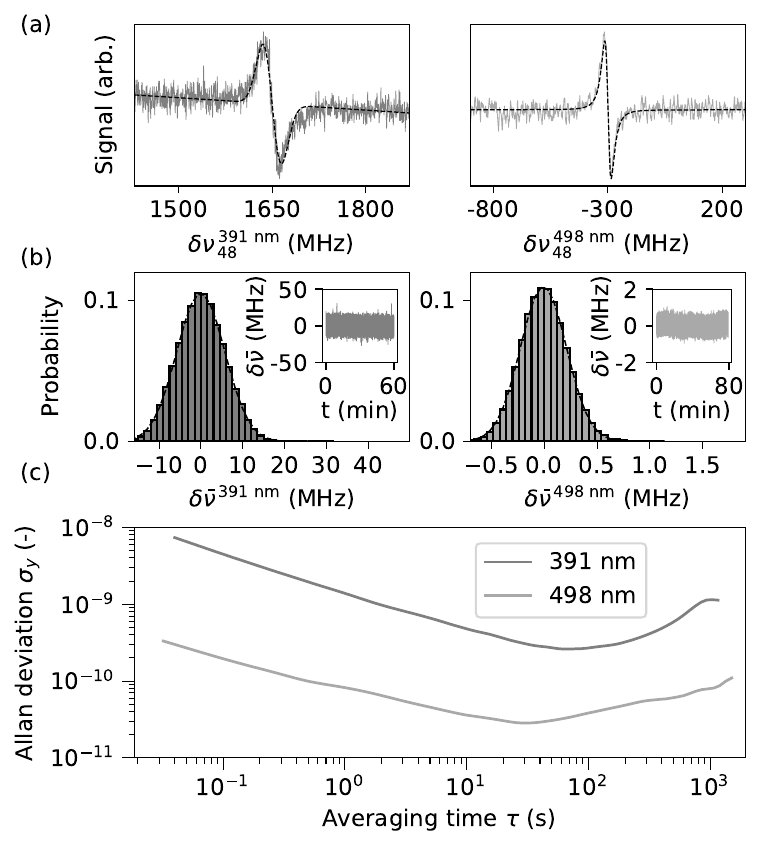}
	\caption{(a) Error signals of the actively locked lasers at \SI{391}{\nm} (dark grey) and \SI{498}{\nm} (light grey), obtained by demodulating MTS signals from the $^{130}\mathrm{Te}_2$ vapor cell with a lock-in amplifier. Before the cell, which was held at \SI{600}{\celsius} for the \SI{391}{\nm} measurement and at \SI{725}{\celsius} for the \SI{498}{\nm} measurement, the \SI{391}{\nm} laser was operated with probe and pump powers of \SI{0.31}{mW} and \SI{0.85}{mW}, respectively, whereas the \SI{498}{nm} laser used \SI{0.15}{mW} (probe) and \SI{21.2}{mW} (pump). The frequency axes are defined relative to the $^{48}\text{Ti}$ optical-pumping line $\mathrm{a^3F_4}\rightarrow\mathrm{y^5D_4^o}$ and the $^{48}\text{Ti}$ laser-cooling transition $\mathrm{a^3F_4}\rightarrow\mathrm{y^5G_6^o}$, respectively. Both signals are fitted to the derivative of a Lorentzian plus a linear background (black, dashed lines). (b) Histograms of frequency deviations from the mean of each actively locked laser, fitted to a Gaussian (black, dashed line). The insets show the corresponding long-term frequency scans with respect to the mean laser frequency. (c) Allan deviation of the actively locked lasers, showing their fractional frequency stability as a function of averaging time. See appendix~\ref{sec:laser-lock-characterization} for further details on the calculation of the Allan deviation.}
\label{fig: allan_dev}
\end{figure}

Fig.~\ref{fig: allan_dev}(c) plots the Allan deviation of the frequency of the locked lasers. The data reveals that the \SI{498}{\nm} lock provides better fractional frequency stability than the \SI{391}{\nm} lock over the full range of averaging times. At short averaging times, both traces follow the expected $\sigma_y\propto\tau^{-1/2}$ scaling of white-frequency noise, consistent with noise in the lock signal dominating the short-term stability. Each curve reaches a clear minimum and then turns upward at long averaging times, indicating the onset of slower correlated noise processes and drift. The \SI{498}{\nm} lock reaches a lower minimum than the \SI{391}{\nm} lock, consistent with higher effective SNR and reduced sensitivity to environmental noise and drift. Although the \SI{391}{\nm} lock is less stable overall, its correct short-term scaling and well-defined minimum suggest that the remaining limitation is primarily technical rather than fundamental.

\section{Conclusion}

In summary, we have demonstrated that \ce{^{130}Te_2} provides a set of reliable frequency references for laser stabilization in the ultraviolet and visible regimes, in particular at lower optical wavelengths than the previously observed wavelength limit of \SI{410}{\nm} \cite{te2_laser_lock_again, te2_mini_spectrum, tellurium_atlas_original}. We demonstrate that these frequency references provide convenient locking points for lasers used to drive atomic resonances in Ti, relevant to optical pumping and laser cooling. Numerous other applications in the same wavelength range can also be pursued. 
\begin{backmatter}
\bmsection{Funding}
This material is based upon work supported by the U.S. Department of Energy, Office of Science, National Quantum Information Science Research Centers, Quantum Systems Accelerator. Additional support is acknowledged from the ONR (Grants No.~N00014-20-1-2513 and No.~N00014-22-1-2280), the ARO (Grants No.~W911NF2010266 and No.~W911NF2310244), and the NSF (PHY-2012068 and the QLCI program through Grant No.~OMA-2016245). L.C.G. acknowledges funding from the UAB Exchange Programme, the University of California Education Abroad Program (UCEAP) and the MOBINT-MIF Scholarship from AGAUR.
\bmsection{Acknowledgment}
We thank Professor Amanda Ross for insightful discussions regarding the band assignment of ditellurium lines in the spectroscopy. 
\bmsection{Disclosures}
The authors declare no conflicts of interest.
\bmsection{Data Availability}
Data underlying the results presented in this paper are not publicly available at this time but may be obtained from the authors upon reasonable request.
\bmsection{Supplemental document}
See supplement for supporting content.
\end{backmatter}

\ifarxiv
    \putbib   % main-text bibliography only
    \end{bibunit}
    \clearpage %clears the page so it starts on a new page
    \appendix
    
    \begin{bibunit}
    
    \setcounter{section}{0}
    \setcounter{figure}{0}
    \setcounter{table}{0}
    \setcounter{equation}{0}
    
    \renewcommand{\thesection}{S\arabic{section}}
    \renewcommand{\thefigure}{S\arabic{figure}}
    \renewcommand{\thetable}{S\arabic{table}}
    \renewcommand{\theequation}{S\arabic{equation}}
\else
  \bibliography{references}

\begin{thebibliography}{10}

\bibitem{iodine_atlas}
S.~Gerstenkorn and P.~Luc.
\newblock {\em {Atlas du spectre d'absorption de la molecule d'iode 14800-20000 cm$^{-1}$}}.
\newblock 1978.

\bibitem{CO_reference}
Shailendhar Saraf, Paul Berceau, Alberto Stochino, Robert Byer, and John Lipa.
\newblock Molecular frequency reference at 1.56$\mu$m using a {$^{12}$C$^{16}$O} overtone transition with the noise-immune cavity-enhanced optical heterodyne molecular spectroscopy method.
\newblock {\em Opt. Lett.}, 41(10):2189--2192, May 2016.

\bibitem{acetylene_reference}
Petr Balling, Marc Fischer, Philipp Kubina, and Ronald Holzwarth.
\newblock Absolute frequency measurement of wavelength standard at 1542 nm: acetylene stabilized {DFB} laser.
\newblock {\em Opt. Express}, 13(23):9196--9201, Nov 2005.

\bibitem{HF_reference}
Shizuo Yamaguchi and Masao Suzuki.
\newblock Frequency locking of an {InGaAsP} semiconductor laser to the first overtone vibration‐rotation lines of hydrogen fluoride.
\newblock {\em Applied Physics Letters}, 41(11):1034--1036, 12 1982.

\bibitem{jun_ye_iodine}
Wang-Yau Cheng, Lisheng Chen, Tai~Hyun Yoon, John~L. Hall, and Jun Ye.
\newblock Sub-doppler molecular-iodine transitions near the dissociation limit (\si{523-498}{nm}).
\newblock {\em Opt. Lett.}, 27(8):571--573, Apr 2002.

\bibitem{barrow_XB_reference}
R.~F. Barrow and R.~P. du~Parcq.
\newblock Rotational analysis of the $\mathrm{A}\,0_\mathrm{u}^+$, $\mathrm{B}\,0_\mathrm{u}^+$-$\mathrm{X}\,0_\mathrm{g}^+$ systems of gaseous {Te$_2$}.
\newblock {\em Proceedings of the Royal Society of London. A. Mathematical and Physical Sciences}, 327(1569):279--287, March 1972.

\bibitem{calcium_tellurium_423}
Jennifer Taylor, Bryan Hemingway, James Hanssen, Thomas~B. Swanson, and Steven Peil.
\newblock Vapor-cell frequency reference for short-wavelength transitions in neutral calcium.
\newblock {\em J. Opt. Soc. Am. B}, 35(7):1557--1562, Jul 2018.

\bibitem{strontium_tellurium_461}
T.~G. Akin, Bryan Hemingway, and Steven Peil.
\newblock Tellurium spectrometer for {{$^{1}\mathrm{S}_{0}$--$^{1}\mathrm{P}_{1}$}} transitions in strontium and other alkaline-earth atoms.
\newblock {\em Review of Scientific Instruments}, 93(5):053002, 05 2022.

\bibitem{barium_ion_tellurium_455}
Tarun Dutta, Debashis~De Munshi, and Manas Mukherjee.
\newblock Absolute {Te$_2$} reference for barium ion at 455.4 nm.
\newblock {\em J. Opt. Soc. Am. B}, 33(6):1177--1181, Jun 2016.

\bibitem{calcium_ion_trapping}
D.~M. Lucas, A.~Ramos, J.~P. Home, M.~J. McDonnell, S.~Nakayama, J.-P. Stacey, S.~C. Webster, D.~N. Stacey, and A.~M. Steane.
\newblock Isotope-selective photoionization for calcium ion trapping.
\newblock {\em Phys. Rev. A}, 69:012711, Jan 2004.

\bibitem{ytterbium_ionization}
M.~Johanning, A.~Braun, D.~Eiteneuer, Chr. Paape, Chr. Balzer, W.~Neuhauser, and Chr. Wunderlich.
\newblock Resonance-enhanced isotope-selective photoionization of {YbI} for ion trap loading.
\newblock {\em Applied Physics B}, 103:327--338, 2011.

\bibitem{calcium_397_state_detection}
F.~Schmidt-Kaler, S.~Gulde, M.~Riebe, T.~Deuschle, A.~Kreuter, G.~Lancaster, C.~Becher, J.~Eschner, H.~Häffner, and R.~Blatt.
\newblock The coherence of qubits based on single {Ca}+ ions.
\newblock {\em Journal of Physics B: Atomic, Molecular and Optical Physics}, 36(3):623, January 2003.

\bibitem{yb_flourescence}
Alec Jenkins, Joanna~W. Lis, Aruku Senoo, William~F. McGrew, and Adam~M. Kaufman.
\newblock Ytterbium nuclear-spin qubits in an optical tweezer array.
\newblock {\em Physical Review X}, 12(2):021027, 2022.

\bibitem{te2_laser_lock_again}
I.S. Burns, J.~Hult, and C.F. Kaminski.
\newblock Use of {$^{130}\mathrm{Te}_2$} for frequency referencing and active stabilisation of a violet extended cavity diode laser.
\newblock {\em Spectrochimica Acta Part A: Molecular and Biomolecular Spectroscopy}, 63(5):905--909, April 2006.

\bibitem{te2_mini_spectrum}
Amanda~J. Ross and Joseph~M. Cardon.
\newblock Te$_2$ absorption spectrum from 19000 to 24000 cm$^{-1}$.
\newblock {\em Journal of Molecular Spectroscopy}, 384:111589, 2022.

\bibitem{titanium_MOT}
Scott Eustice, Jackson Schrott, Anke St\"oltzel, Julian Wolf, Diego Novoa, Kayleigh Cassella, and Dan~M. Stamper-Kurn.
\newblock Magneto-optical trap of titanium atoms.
\newblock {\em Phys. Rev. Res.}, 7:023025, Apr 2025.

\bibitem{eustice_optical_2023}
Scott Eustice, Dmytro Filin, Jackson Schrott, Sergey Porsev, Charles Cheung, Diego Novoa, Dan~M. Stamper-Kurn, and Marianna~S. Safronova.
\newblock Supplemental material for “optical telecommunications-band clock based on neutral titanium atoms”.
\newblock {\em Physical Review A}, 107(5):L051102, May 2023.

\bibitem{tellurium_atlas_original}
P~Luc and J~Cariou.
\newblock {\em Atlas du spectre d'absorption de la molecule de tellure}.
\newblock Laboratoire Aime-Cotton CNRS II, Orsay, France, 1980.

\bibitem{MTS_paper}
Jon~H. Shirley.
\newblock Modulation transfer processes in optical heterodyne saturation spectroscopy.
\newblock {\em Opt. Lett.}, 7(11):537--539, Nov 1982.

\bibitem{e8_isotope_shifts_2021}
Andrew~O. Neely, Kayleigh Cassella, Scott Eustice, A~A, and Dan~M. Stamper-Kurn.
\newblock Isotope shifts in the metastable {${a}^{5}\mathrm{F}$} and excited {${y}^{5}{\mathrm{G}}^{o}$} terms of atomic titanium.
\newblock {\em Phys. Rev. A}, 103:032818, Mar 2021.

\bibitem{pressure_vs_temp_pressure_broadening}
Arnolds Ubelis.
\newblock Temperature dependence of the saturated vapor pressure of tellurium.
\newblock {\em Journal of Engineering Physics}, 42:309--315, 03 1982.

\bibitem{Residual_doppler_broadening}
Isao Hirano.
\newblock Effect of intersection angle between saturation and probe beams on saturated absorption spectra of the {Cs}-{D2} line.
\newblock {\em Journal of Quantitative Spectroscopy and Radiative Transfer}, 40(4):531--537, 1988.

\bibitem{band_head_data}
B.L. Jha, K.V. Subbaram, and D.Ramachandra Rao.
\newblock Electronic spectra of {$^{130}\mathrm{Te}_2$} and {$^{128}\mathrm{Te}_2$}.
\newblock {\em Journal of Molecular Spectroscopy}, 32(3):383--397, December 1969.

\end{thebibliography}


\begin{thebibliography}{1}

\bibitem{band_head_data}
B.L. Jha, K.V. Subbaram, and D.Ramachandra Rao.
\newblock Electronic spectra of {$^{130}\mathrm{Te}_2$} and {$^{128}\mathrm{Te}_2$}.
\newblock {\em Journal of Molecular Spectroscopy}, 32(3):383--397, December 1969.

\bibitem{barrow_XB_reference}
R.~F. Barrow and R.~P. du~Parcq.
\newblock Rotational analysis of the $\mathrm{A}\,0_\mathrm{u}^+$, $\mathrm{B}\,0_\mathrm{u}^+$-$\mathrm{X}\,0_\mathrm{g}^+$ systems of gaseous {Te$_2$}.
\newblock {\em Proceedings of the Royal Society of London. A. Mathematical and Physical Sciences}, 327(1569):279--287, March 1972.

\end{thebibliography}
\fi

\section*{Supplementary Material}

\title{\texorpdfstring{\ce{^{130}Te_2}}{130Te2} spectroscopic reference for neutral Ti lines at 391 nm and 498 nm: supplemental document}

\section{Temperature dependence of \texorpdfstring{\ce{^{130}Te_2}}{130Te2} MTS signals}
\label{sec:temp-scans}

\begin{figure}[!htb]
    \centering
    \includegraphics[width=\linewidth]{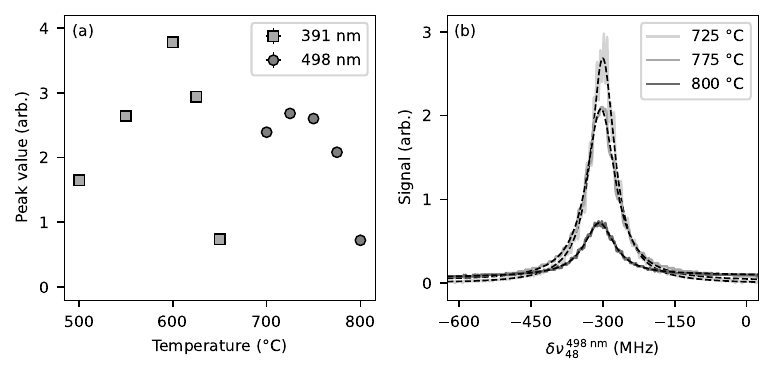}
    \caption{Temperature dependence of the \ce{^{130}Te_2} MTS signal amplitude used for laser spectroscopy. The spectra were recorded while varying the furnace temperature of the $^{130}\mathrm{Te}_2$ vapor cell. Panel (a) shows the \ce{^{130}Te_2} MTS peaks, in arbitrary units, at 391 and \SI{498}{nm} optical wavelengths. The strongest signal, and therefore the highest error-signal SNR, is obtained at \SI{600}{\celsius} for 391 nm and \SI{725}{\celsius} for \SI{498}{nm}. Panel (b) shows representative traces of the feature used to lock the laser to the Ti cooling transition at $\lambda = \SI{498}{\nano\meter}$. The signal amplitude reaches a maximum at \SI{725}{\celsius}, which is therefore chosen as the operating temperature for locking to this resonance.}

    \label{fig:temp_scans}
\end{figure}
\par
To maximize the SNR of the \SI{391}{nm} and \SI{498}{nm} optical transitions, we optimized the temperature of the vapor cell in each case. After optimizing at \SI{391}{nm}, we observe that among the 36 Doppler-broadened absorption features, 21 also exhibit a Doppler-free resonance. At higher temperatures, the pump beam intensity is strongly attenuated as it passes through the cell, reducing the MTS contrast for strong lines, while weaker lines may gain MTS contrast because the higher absorber density compensates for sub-saturation fields. Certain weaker lines, however, may have too low of a transition strength for our laser to sufficiently depopulate the ground state and thus produce an MTS signal. For the \qty{256}{\GHz} scan, we optimize the temperature once for the \SI{765.6685}{\THz} resonance used for locking (Sec.~\ref{sec:identification-locking}), accepting that not all of the transitions will have simultaneously optimal MTS contrast.

\section{Fitting of observed \texorpdfstring{\ce{^{130}Te_2}}{130Te2} lines and molecular line assignment}
\label{sec:line-assignment}

Since the dominant form of line-broadening for Doppler-broadened spectroscopy is largely Doppler broadening, we fit each observed line to a Gaussian. MTS spectra are fit to Lorentzian profiles. We also fit the MTS data to Voigt profiles to include both homogeneous and inhomogeneous broadening mechanisms, but found that the data was better fit by a Lorentzian profile. In addition to the wavemeter-induced \SI{30}{\MHz} uncertainty in the absolute frequency of all lines, we deduce the $1\sigma$ statistical uncertainty in fit parameters through diagonal elements in the covariance matrix of the nonlinear least-squares fit. The covariance matrix is estimated from the inverse curvature of the least-squares objective evaluated at the best-fit solution. The Doppler-broadened features are additionally fit with a linear offset per scan to account for an observed etalon with $\sim$\SI{50}{\GHz} spectral range. Fitting each scan with a broad sinusoidal offset did not yield good fits due to the inter-scan offsets imposed by conversion from photodiode voltage to absolute units of absorbance. We tabulate these fitted line centers and uncertainties in Table~\ref{tab:doppler_vs_MTS_sorted}.

From Sec.~\ref{sec: broadband spectroscopy}, we can deduce the electronic subsystem to which the ultraviolet spectroscopy pertains. Furthermore, from prior band-head data~\cite{band_head_data}, we determine that the observed lines reside within the following vibrational transitions: $\nu=(28,27,26,25,24)\rightarrow 0$ and $(27,26)\rightarrow 1$. Uncertainties in the rovibrational constants limit a more precise line-by-line assignment. The quoted uncertainty in the rotational constant $B_\nu$ ($\pm 5\times 10^{-6}\,\text{cm}^{-1}$) yields an energy uncertainty that scales with $(J^2+J)$ and corresponds to roughly \SI{12}{\GHz} at high $J$, while the $\pm 5\times 10^{-2}\,\text{cm}^{-1}$ uncertainty in the vibrational constant $\omega_\nu$ contributes an additional \SI{8}{\GHz}. Moreover, above $\nu_B=19$, the vibrational bands exhibit significant perturbations from a linear model $\Delta G(\nu+1/2)$, thus preventing a more precise vibrational band assignment \cite{barrow_XB_reference, band_head_data}.

\section{Laser lock characterization}
\label{sec:laser-lock-characterization}

We quantify the frequency stability of the frequency-stabilized lasers with the Allan deviation. The wavemeter records wavelength measurements of the actively locked lasers at \SI{391}{nm} and \SI{498}{nm} at intervals of $\Delta t=\SI{40}{\ms}$ and $\Delta t=\SI{30}{\ms}$, respectively. We convert the recorded wavelengths into optical frequencies and then into fractional frequency deviations,
\begin{equation}
    y_i = \frac{\nu_i-\bar{\nu}}{\bar{\nu}},
\end{equation}
where $\nu_i$ is the instantaneous frequency corresponding to the $i^{\mathrm{th}}$ measurement, and $\bar{\nu}$ is the mean frequency over the full record.

Next, we define 50 logarithmically spaced averaging factors $m\in\mathbb{Z}^+$, ranging from $1$ to $m_{\max}=\lfloor N/2\rfloor$, where $N$ is the total number of measurements. For each $m$, the corresponding averaging time is $\tau=m\Delta t$, and the averaged fractional frequency is defined as
\begin{equation}
    \bar{y}_i(\tau)=\frac{1}{m}\sum_{k=0}^{m-1} y_{i+k},
\end{equation}
where $i\in\{1,2,\dots, N-m+1\}$ denotes the index of the first measurement included in the average. Since $i$ increases by one from one average to the next, these averages are formed from overlapping sets of measurements.

\begingroup
\footnotesize
\setlength{\tabcolsep}{4pt}
\renewcommand{\arraystretch}{0.85} 

\begin{table}[htbp]
\centering
\begin{tabular}{c|c|c||c|c}
\multicolumn{3}{c||}{\textbf{Doppler Broadened}} & \multicolumn{2}{c}{\textbf{MTS}} \\
\hline
\textbf{Center} & \textbf{FWHM} & \textbf{Absorb.} & \textbf{Center} & \textbf{FWHM} \\
\textbf{(GHz)}  & \textbf{(MHz)} &                & \textbf{(GHz)}  & \textbf{(MHz)} \\
\hline
-103.375(36)   & --      & 0.002(1)   &                &            \\
-91.826(37)    & --      & 0.001(1)    &                &            \\
-85.220(6)     & --      & 0.004(3)    &                &            \\
-80.735(13)    & 1084.4(3.5)  & 0.006(1)   & -80.759(11)    & --    \\
-77.188(10)    & 970.5(1.2)   & 0.181(3)   & -77.158(10)    & 88.00(61)  \\
-70.456(13)    & 953.6(3.4)   & 0.061(4)   & -70.432(11)    & 57.0(1.3)  \\
-68.410(18)    & --     & 0.006(2)    &                &            \\
-60.809(26)    & --     & 0.006(2)    &                &            \\
-56.701(24)    & --     & 0.007(2)  &                &            \\
-52.812(10)    & 1185(18)     & 0.007(2)   &                &            \\
-44.269(9)     & 1425(17)     & 0.009(2)   &                &            \\
-40.813(10)    & 1028.5(1.3)  & 0.099(1)    & -40.768(10)    & 35.00(70)  \\
-35.177(11)    & --      & 0.0036(9)   &                &            \\
-29.506(3)     & 1094.1(7.6)  & 0.027(2)    & -29.483(10)    & 165.0(3.5) \\
-27.767(2)     & 951.4(2.0)   & 0.094(2)    & -27.746(10)    & 72.00(55)  \\
-20.264(4)     & 1049(13)     & 0.012(2)     &                &            \\
-18.177(10)    & 962.3(2.0)   & 0.036(3)    & -18.180(10)    & 39.00(2.0) \\
-15.609(10)    & 989.54(0.67) & 0.181(4)    & -15.585(10)    & 80.00(0.91)\\
-6.126(6)      & 1120(20)     & 0.004(1)    &                &            \\
1.659(10)     & 1009.7(2.7)  & 0.128(2)    & 1.680(10)     & 63.00(0.70)\\
3.338(10)     & 1216.7(4.7)  & 0.009(2)    &                &            \\
6.775(10)     & 883.8(1.4)   & 0.015(2)    & 6.692(10)     & --   \\
12.819(9)     & 1132(23)     & 0.008(2)    &                &            \\
31.897(10)    & 982.7(1.4)   & 0.177(4)    & 31.920(10)    & 78.00(52)  \\
42.249(10)    & 997.9(1.8)   & 0.093(4)    & 42.254(10)    & 43.00(54)  \\
47.395(10)    & 1000.4(1.4)  & 0.111(4)    & 47.421(10)    & 33.00(66)  \\
52.380(10)    & 880(11)      & 0.020(3)    & 52.408(10)    & --  \\
55.981(10)    & 994.2(2.1)   & 0.100(4)   & 55.994(10)    & 51.00(45)  \\
86.626(10)    & 1039.532(34) & 0.018(4)   &                &            \\
91.764(10)    & 983.3(1.5)   & 0.191(5)    & 91.791(10)    & 75.00(67)  \\
101.090(10)   & 1090.281(23) & 0.007(3)    &                &            \\
107.613(10)   & 1016.110(17) & 0.010(3)   &                &            \\
122.321(10)   & 973.2(0.91)  & 0.092(1)   & 122.337(10)   & 65.00(91)  \\
123.844(10)   & 980.1(1.2)   & 0.069(1)  & 123.858(10)   & 81.00(92)  \\
133.044(10)   & 1027.474(3.1)& 0.036(3)   & 133.164(29)   & 250.00(85) \\
135.958(10)   & 967.2(1.1)   & 0.096(4)    & 135.984(10)   & 81.00(2)   \\
138.384(10)   & 993.92(0.63) & 0.173(5)   & 138.407(10)   & 117.00(1.2)\\
\hline
\end{tabular}
\caption{Fits of Doppler-broadened and MTS peaks, with statistical uncertainties. The frequencies shown are relative to the $\mathrm{a^3F_4\rightarrow y^5D_4^o}$ optical pumping line at \SI[parse-numbers=false]{765.666836(3)_{\mathrm{stat}}(30)_{\mathrm{sys}}}{\THz} in atomic $^{48}$Ti. Some resonances were not seen in MTS, and their corresponding entries are left blank. For particularly weak resonances, we were unable to reliably determine the resonance width, potentially due to the presence of multiple overlapping resonances. The FWHM of these features is indicated by a dash (--). The oscilloscope analog-to-digital converter imparts a $0.001$ uncertainty on all peak absorbances, which is greater than the calculated fitting error and thus is reported for all values. At \SI{600}{\celsius}, the expected FWHM resulting from Doppler broadening is \SI{1006}{\MHz}, which closely follows the reported widths. Because the amplitudes of the MTS signals depend strongly on the pump power, cell temperature, and system alignment, they are not inherent quantities of the lines. As such, MTS signal amplitude is not tabulated.}
\label{tab:doppler_vs_MTS_sorted}
\end{table}

\endgroup

For each averaging time $\tau$, the Allan deviation is calculated from the mean squared difference between averaged fractional frequencies separated by $m$ measurements,
\begin{equation}
    \sigma_y(\tau)=\sqrt{\frac{1}{2}\left\langle\left(\bar{y}_{i+m}(\tau)-\bar{y}_i(\tau)\right)^2\right\rangle}
    =\sqrt{\frac{1}{2L}\sum_{i=1}^{L}\left(\bar{y}_{i+m}(\tau)-\bar{y}_i(\tau)\right)^2},
\end{equation}
where $L=N-2m+1$ is the number of squared differences included in the average. Each difference compares two adjacent, non-overlapping groups of $m$ measurements, corresponding to a duration $\tau$. Repeating this procedure for all values of $m$ yields the Allan deviation as a function of averaging time.

In Fig.~\ref{fig: allan_dev}(c), for $\tau\leq\SI{3}{s}$, both locks exhibit $\sigma_y\propto\tau^{-1/2}$ behavior characteristic of white-frequency noise, indicating that random frequency fluctuations from detection noise (including shot noise) in the locking system limit the short-term stability. The Allan deviation minima occur at $\sigma_y=2.8\times10^{-11}$ at $\tau=\SI{30}{s}$ for the \SI{498}{\nm} lock and $\sigma_y=2.6\times10^{-10}$ at $\tau=\SI{67}{s}$ for the \SI{391}{\nm} lock. The earlier and lower minimum for \SI{498}{\nm} indicates higher effective SNR and reduced technical or environmental coupling. For $\tau\gtrsim10^2\,\text{s}$, both curves turn upward, consistent with random-walk FM and frequency drift, though some portion of the observed uptick may originate from the wavemeter itself through slow drift and infrequent recalibration to our \SI{780}{\nm} reference light. The \SI{391}{\nm} lock shows a stronger rise, indicating larger long-term instabilities. Despite its higher Allan deviation, the \SI{391}{\nm} lock exhibits the expected noise scaling and a clear stability minimum, suggesting that the present performance gap is due to technical imperfections rather than a fundamental limit in the potential lock stability.

\putbib   % appendix/supplement bibliography only
\end{bibunit}

\end{document}